\documentclass[twocolumn,journal]{IEEEtran}
\usepackage{graphicx}
\usepackage{dcolumn}
\usepackage[noadjust]{cite}
\usepackage{amsmath}
\newcommand{\be}{\begin{eqnarray}}
\newcommand{\ee}{\end{eqnarray}}
\newcommand{\ket}[1]{|#1\rangle}
\newcommand{\mi}{\mathrm{i}}
\newcommand{\bra}[1]{\langle#1|}

\begin{document}
\title{Double Electromagnetically Induced Transparency in a Tripod-type Atom System}
\author{Xiao Li,
    Yu Liu,
    Hong Guo$^\ast$ %
\thanks{Xiao Li, Yu Liu and Hong Guo are with CREAM Group, School of Electronics Engineering and Computer
Science (EECS), Peking University, and Xiao Li is also with School of Earth and Space Science,
Peking University, Beijing 100871, P. R. China. }%
\thanks{$\ast$Author to whom the correspondence should be
addressed. Phone: +86-10-6275-7035, Fax: +86-10-6275-3208, E-mail: hongguo@pku.edu.cn.}}

\maketitle
\begin{abstract}
The electromagnetically induced transparency (EIT) phenomenon in a four level atomic system with
tripod configuration is studied. The results show that this configuration is equivalent to the
combination of two single three-level $\Lambda$ configurations, which, under certain conditions,
results in the so-called double-EIT (DEIT) phenomenon. The properties of the double transparency 
windows for DEIT are discussed in detail and the possible experimental scheme is proposed.
\end{abstract}

\section{Introduction}
\IEEEPARstart{E}{lectromagnetically} induced transparency (EIT), as one of the quantum coherence
and interference (QCI) phenomena, is an important effect induced by the interaction between laser
beams and atom ensembles under two-photon resonance
condition\,\cite{harris1997pt}-\cite{fleischhauer2005rmp}. EIT in three-level systems, such as the
typical $\Lambda$-type system, has been extensively studied both theoretically and
experimentally\,\cite{fulton1995pra}. Its direct results, such as subluminal\,\cite{hau1999nature}
and superluminal\,\cite{wang2001nature} light propagations, have already been demonstrated
experimentally. After that, with the model of dark state
polariton\,\cite{fleischhauer2000prl}-\cite{lukin2001nature}, the EIT-based light storage is
theoretically proposed and then experimentally realized\,\cite{liu2001nature}-\cite{han2005pla}.

For some types of four-level atomic configurations, similar analysis also shows some interesting
phenomena, such as the interchange between subluminal and superluminal propagation
\cite{gonzalo2005pra} and double EIT phenomenon \cite{knight2002pra}. However, for most of the
four-level systems, there is hardly ideal dark states and so brings serious limitations to their
applications. Fortunately, as we will show in this paper, the four-level tripod-type atom system
does yield an ideal dark state, which may greatly improve the coherence performances of current
four-level atomic systems.

\begin{figure}[!htb]
  \centering
  \includegraphics[scale=0.4]{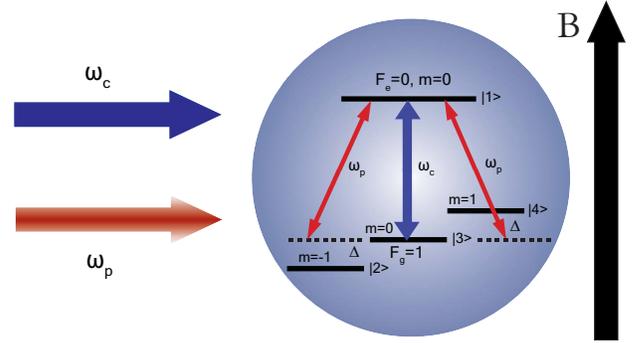}
  \caption{(color online) The configuration of the tripod-type four-level atom.
  $\ket{1}$ is the excited state with the damping rates $\Gamma_{12}$,
  $\Gamma_{13}$, and $\Gamma_{14}$. $\ket{2}$, $\ket{3}$, $\ket{4}$ are the
  ground states split by an external magnetic field, while $\Delta$ is the
  splitting gap. Levels $\ket{1}$ and $\ket{2}$, $\ket{1}$
  and $\ket{4}$ are coupled by the left and right polarized components of a
  weak probing light $\Omega_p$, respectively, while levels $\ket{1}$ and
  $\ket{3}$ are coupled by a strong light $\Omega_c$.}
  \label{configuration}
\end{figure}

In Fig.\,\ref{configuration}, we show the schematic setup of a four-level tripod-type atomic
system. A linearly polarized ($\pi$) light is served as the coupling light while the probing light
is composed of left and right polarized components $\sigma_{\pm}$ with the same frequency
$\omega_p$ and the same amplitude. In this configuration, the probing light is traveling in the
direction perpendicular to both the polarization of the coupling light and the direction of the
magnetic field, which is used to split the ground level into three Zeeman sub-levels
($m_{\textrm{F}} = -1,0,+1$). Thus, this system is equivalent to the combination of two simple
$\Lambda$-type systems coupled by a common light.

  It has been shown \cite{yinggu2003pra} that the coherence situation
 of this configuration can be altered by adjusting the the Rabi frequency of the coupling
 light $\Omega_c$ and the Zeeman splitting $\Delta$. The double EIT phenomenon can thus be
 modulated by the Rabi frequency $\Omega_c$ and thereby the frequency differences between the
 coupling and the probing light, i.e.,  $\delta = \omega_p-\omega_c$.
 However, there are following issues left to be explored:
 1) a systematic discussion of the role of each parameters in
 this configuration, 2) the impact of the non-radiative damping on the coherence of the system,
 and 3) the conditions required for experimental realization. All of the above issues will be
 discussed in this paper. We will first give numerical solutions to the equation of motion for a
 tripod-type four-level atomic system and discuss its absorption and dispersion properties.
 Then, explicit expressions will be given to show that the four-level tripod-type system is indeed a
 combination of two three-level $\Lambda$-type systems. Following that, we will show how to
 make this system an optimal EIT system by adjusting related parameters and then, we will discuss
 the experimental possibilities.

\section{Absorption and dispersion for a tripod-type
 four-level atomic system}

In the four-level atomic system shown in Fig.\,\ref{configuration}, the Hamiltonian under the
rotating wave frame can be written as
    \begin{align}
      \hat{H} &= \hbar (\omega_p-\omega_{12}) \ket{2}\bra{2}+ \hbar(\omega_c-\omega_{13})
      \ket{3}\bra{3} \notag\\
      &+ \hbar (\omega_p-\omega_{14}) \ket{4}\bra{4}\notag\\
      &-\frac {\hbar}{2} \left(\Omega_p
      \ket{1}\bra{2}+\Omega_c\ket{1}\bra{3}+\Omega_p\ket{1}\bra{4}+\mathrm{H.c.}\right),
    \end{align}
where $\Omega_p$ and $\Omega_c$ are the Rabi frequencies and $\omega_{ij} = \omega_i-\omega_j$ is
the central frequency between Zeeman sub-levels $\ket{i}$ and $\ket{j}$.

A general expression of the eigenstates of the Hamiltonian is very complicated. However, when the
Zeeman splitting is equal to the frequency difference between the two lights, an ideal dark state,
whose eigenvalue is zero, emerges, with the expression:
    \begin{align}
      \ket{\Psi^0} = -\ket{2}+\ket{4},\label{darkstate}
    \end{align}
while the other three eigenstates are
    \begin{align*}
      \ket{\Psi_{i}} = -\frac{2\lambda_i}{\hbar\Omega_p}\ket{1}
      +\ket{2}-2\frac{\hbar^2\Omega_p^2-2\lambda_i^2}{\hbar^2\Omega_c\Omega_p}\ket{3}+\ket{4},
    \end{align*}
with eigenvalues $\lambda_i\,(i = 1,2,3)$ satisfying
    \begin{align*}
      4\lambda^3-8\hbar\delta_c\lambda^2-\hbar^2(\Omega_c^2+2\Omega_p^2)\lambda+4\hbar^3\delta_c\Omega_p^2
      = 0.
    \end{align*}

Next we will take into account the decays of the atomic levels due to radiative and non-radiative
dampings. We base our discussion on the steady state solution of the density matrix equation. For
simplicity, we assume that $\Omega_p$ and $\Omega_c$ are real and the coupling light interacting
with the states $\ket{1}$ and $\ket{3}$ is resonant ($\omega_c = \omega_{13}$). Suppose that all
radiative damping rates are equal, i.e., $\Gamma_{12} = \Gamma_{13} = \Gamma_{14} = \beta\gamma$,
where $\gamma$ is atomic spontaneous emission rate, and so are the nonradiative damping rates,
i.e., $\Gamma_{ij} = \alpha\gamma$, where $i,j = 2,3,4\,(i\neq j)$. Also, we set $\omega_{32} =
\omega_{43} = \Delta\gamma$, $\Omega_p = g_p \gamma$, $\Omega_c = g_c\gamma$, and $\delta =
\omega_p-\omega_c = \delta_c\gamma$ for normalization. Here, $\Gamma_{ij}$ represents the damping
rate from state $\ket{i}$ to $\ket{j}$, $\omega_{ij} = \omega_i-\omega_j$, and $\delta$ is the
frequency difference between the coupling and probing lights. Under the above assumptions, we can
get the steady-state solution for these equations.

\begin{figure}[!htb]
  \centering
  \includegraphics[scale=0.5]{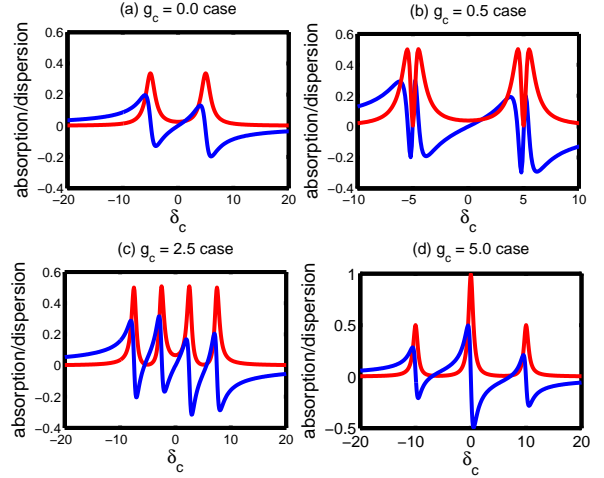}
  \caption{(color online) Absorption and dispersion of the four-level tripod configuration.
  Here the blue curve represents the dispersion, while the red curve represents absorption. The parameters
  are $\Delta = 5.0 $, $\alpha = 0.001$, and $\beta = 0.666$. The differences result from the different
  values of $g_c$.}
  \label{numericalsolution}
\end{figure}

To see the absorption and dispersion characteristics of the probing light, we plot the steady state
solution of $(\rho_{12}+\rho_{14})/g_p$ in Fig.\,\ref{numericalsolution}, where $\Delta = 5.0$,
$\alpha = 0.001$, and $\beta = 0.666$. From the figure, one finds that this four-level atomic
system has double transparency windows, each of which is the typical result of a $\Lambda$-type
system. The concrete discussion on this result will be given in the next section.

The following analytical results will show that the tripod scheme can, under some approximations,
be viewed as a combination of two $\Lambda$-type schemes. Firstly, the expression for
$(\rho_{12}+\rho_{14})/g_p$, which represents the absorption and dispersion characteristics of the
tripod system, can further be  simplified if we neglect the terms with higher orders of $g_p$, as
shown below:
    \begin{align}
      h =\frac{\rho_{12}+\rho_{14}}{g_p}= \frac{\mi \beta}{9\alpha \beta  +
    4\left( 2\alpha  + \beta\right)g_c^2}(h_l+h_r),\label{doublewindow}
     \end{align}
where
    \begin{align*}
      h_l =\frac{3\alpha  \left( 2\alpha  + \mi \Delta  -\mi\delta_c \right)
      +g_c^2\left( \alpha  +2\mi  \Delta - 2\mi\delta_c\right) }
      {g_c^2 - \left( \mi- \Delta  + \delta_c\right)
       \left(2 \mi\,\alpha  - \Delta  + \delta_c\right)},\\
       h_r =\frac{3\alpha  \left( 2\alpha-\mi \Delta  -\mi \delta_c \right)
       +g_c^2\left( \alpha-2\mi\Delta - 2\mi \delta_c\right) }
       {g_c^2 - \left( \mi+ \Delta  + \delta_c\right)
        \left(2 \mi\,\alpha+\Delta  + \delta_c\right)}.
    \end{align*}

For simplicity, the damping rates between the ground states are ignored at the moment and the
validity of this approximation will be discussed at the end of the paper. Then, one has:
    \begin{align}
      h =\frac{\rho_{12}+\rho_{14}}{g_p}=\frac 1 2 \left[ h_0(\delta_c-\Delta)+h_0(\delta_c+\Delta)\right],
      \label{formula}
    \end{align}
with
    \begin{align}
      h_0(x) = \frac{x}{g_c^2- x (\mi+x)}.\label{Equation:hx}
    \end{align}
It is clear from Eq. (\ref{formula}) that the system  is equivalent to a combination of two
symmetric parts. Besides, some important properties of $h_0(x)$ is worth discussing here. We note
that, according to Eq.~\eqref{Equation:hx}, its imaginary part reads
   \begin{align}
     \text{Im}[h_0(x)] = \frac{x^2}{(g_c^2-x^2)^2+x^2},
   \end{align}
One can easily verify that this function has a minimum of zero at $x=0$. Also, it has two peaks at
$x = \pm\,g_c$, respectively. Therefore, the distance between the two peaks is $2g_c$. These
conclusions will be useful in our later discussion. We now go on to show that each part of
Eq.\,\eqref{formula} is exactly the same as that of a $\Lambda$-type scheme.

\begin{figure}[!htb]
  \centering
  \includegraphics[scale=0.4]{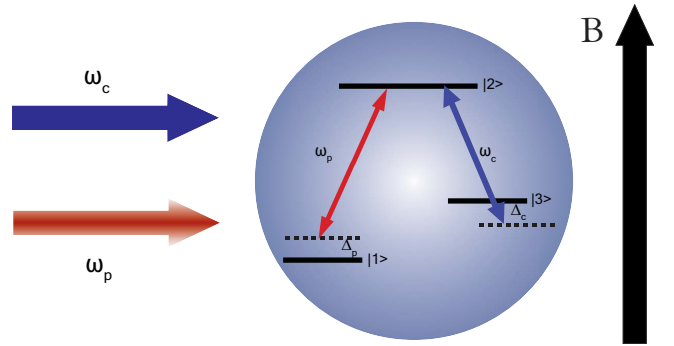}
  \caption{(color online) The configuration of the three level $\Lambda$-type atom.
  $\ket{1}$ and $\ket{3}$ are two ground state sub-levels while $\ket{2}$ is the excited state
  with the damping rates $\Gamma_{21}$ and $\Gamma_{23}$.
  $\Omega_p$ and $\Omega_c$ are the Rabi frequencies of the probing and the coupling light,
  respectively. $\Delta$ is the splitting caused by Zeeman effect.}
  \label{lambdaconfig}
\end{figure}

The standard three level $\Lambda$-type scheme is shown in Fig.\,\ref{lambdaconfig}. Following the
same approach as mentioned above, we first give the steady-state solution of $\rho_{21}$, the real
and imaginary part of which represent the dispersion and the absorption properties of the system,
respectively:
    \begin{align*}
     h_1 =\;& -A/B,
    \end{align*}
 where
    \begin{align*}
 A =&g_c^2\left[ g_c^2 + g_p^2 -
        \left( \Delta  - \delta_c \right) \,\left(\mi  + \Delta  -\delta_c \right)  \right] \,
      \left( \Delta  - \delta_c \right),
      \\
 B =&g_c^2\left[ 3\,g_p^4 \,+%
 \left( \Delta  - \delta_c\right)^2 \left( 1 + \Delta^2 - 2\,\Delta \,\delta_c
           + \delta_c^2 + 4\,g_p^2\right)\right]\\
 +&g_c^4\left[ 3\,g_p^2 - 2\,\left( \Delta  - \delta_c\right)^2 \right]  + g_c^6 +
      g_p^2\,\left[g_p^4 + \left( \Delta  - \delta_c \right)^2\right].
    \end{align*}

\begin{figure}[!htb]
  \centering
  \includegraphics[scale=0.5]{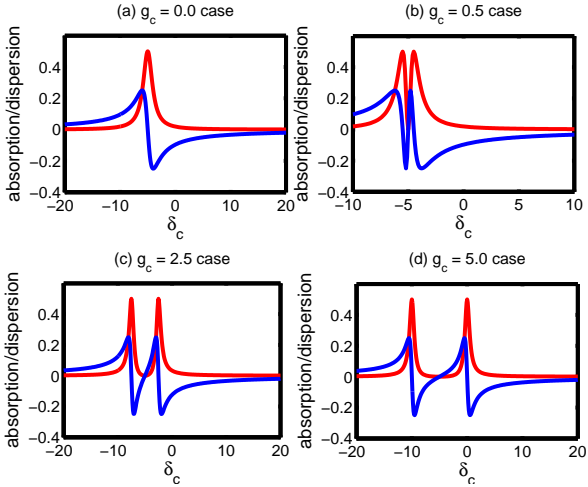}
  \caption{(color online) Dispersion and absorption of the three-level $\Lambda$
  configuration. Likewise, the blue curve represents the dispersion, while the red curve represents the absorption.
  The related parameters are $\Delta = 5.0$, $\alpha = 0.001$, and $\beta = 0.666$. $\Omega_c$ is used to
  modulate the system. From the figures,
  one recognizes that the tripod-type is equivalent to the combination of two $\Lambda$ systems.}
  \label{lambdafig}
\end{figure}

Since the probing light is much weaker than the coupling light, the above expression can be
simplified as:
    \begin{align}
      h_1=\frac{\delta_c-\Delta}{g_c^2 - \left(\delta_c-\Delta\right) \,\left(\mi+\delta_c-\Delta\right)},
    \end{align}
which is exactly the same as $h_0(\delta_c-\Delta)$, the first term of the four-level result [see
Eq. (\ref{formula})]. Here, $\Omega_c = g_c\gamma$ is the Rabi frequency of the coupling light,
$\Delta\gamma = \omega_{31}$ denotes the Zeeman splitting of the ground level, and $\delta_c\gamma
= \omega_p-\omega_c$ is the frequency difference between the coupling and the probing lights.
Fig.\,\ref{lambdafig} shows the dispersion and absorption properties of this three-level system,
which further confirms our conclusion that the tripod-type atomic system can, under the weak
probing light assumption, be viewed as the combination of two independent $\Lambda$-type systems.

\section{The impact of the coupling light and the magnetic field}
In the previous section, we have reached the conclusion that the tripod-type atomic system is
equivalent to the combination of two $\Lambda$-type ones. Here we want to figure out in what way
the two $\Lambda$-type schemes constitute the two transparency windows.

Recalling the properties of $h_0(x)$, we can conclude that $h_0(\delta_c-\Delta)$ has its minimum
at $\delta_c-\Delta=0$, with the two peaks at $\delta_c-\Delta=\pm\,g_c$, respectively. That is to
say, the right transparency window will have a central frequency of $\delta_c = \Delta$ and a width
$2g_c$. This conclusion is also valid for the left transparency window, i.e., it is centered at
$\delta_c = -\Delta$ with the width $2g_c$. The above discussion gives us some hint on how to
construct the two transparency windows. If we choose to fix the external magnetic field and scan
the frequency of the probing light, the two transparency windows are expected to emerge when the
frequency difference between the probing light and the pumping light satisfies $\delta_c = \pm
\Delta$, respectively.

\begin{figure}[!htb]
  \centering
  \includegraphics[scale=0.5]{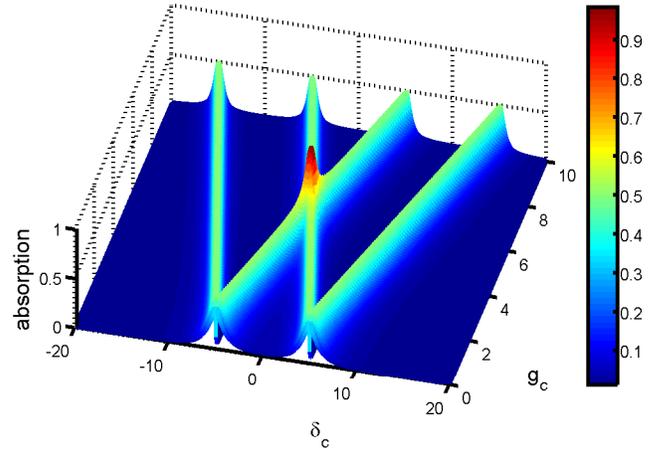}
  \caption{(color online) The impact of the coupling light. Here we keep the Zeeman splitting
  $\Delta = 5$ while the Rabi frequency of the
  coupling light $g_c$ is modulated. The color represents
  the magnitude of the absorption strength, as shown in the bar. The absorption and dispersion
   under four typical values of $g_c$ have been shown in Fig.~\ref{numericalsolution}.}
  \label{Fig:modulationlight}
\end{figure}

The Rabi frequency of the coupling light $\Omega_c = g_c\gamma$ is of interest here. We want to
show that it should not be too large in order to obtain satisfied transparency windows. The first
reason is very apparent, since a narrow transparency window is always desirable for obtaining slow
light and thereby the light storage. The second reason is not so close at hand: We will show this
point in Fig.\,\ref{Fig:modulationlight}, where the Zeeman splitting is kept constant $\Delta = 5$.
One can see that with the increase of $g_c$, the width of the two transparency windows becomes ever
larger. When $g_c>\Delta$, however, the two windows begin to have overlap, which significantly
alters the absorption properties of the system. In this case, the two windows will be centered at
$\pm\,g_c$, with the width of $2\Delta$. Given the above two reasons, we conclude that when using
the tripod-type scheme, the Rabi frequency of the coupling light is preferable to be less than the
Zeeman splitting. There is another interesting result in Fig.\,\ref{Fig:modulationlight}. When
$g_c$ equals to $\Delta$, we observe the so-called EIA phenomenon\,\cite{lezama1999pra}, in which
the absorption of the system is doubled at $\delta_c = 0$.

\begin{figure}[!htb]
  \centering
  \includegraphics[scale=0.5]{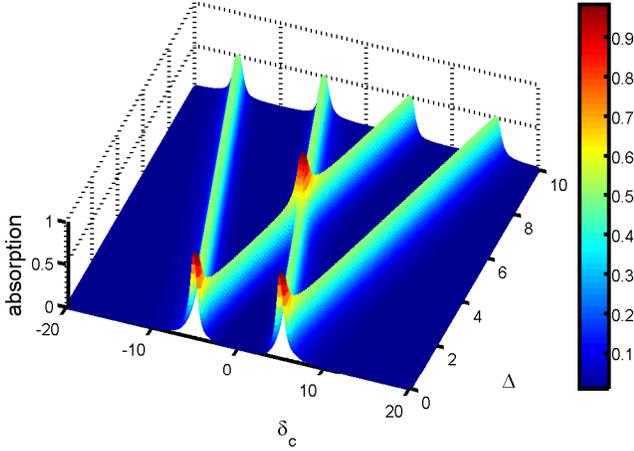}
  \caption{(color online) The impact of the magnetic field. Here we keep the Rabi frequency of the
  coupling light to be $g_c = 5$ while the Zeeman splitting $\Delta$ is modulated. The color represents
  the magnitude of the absorption, as shown in the bar. Fig.\,\ref{Fig:varydelta} shows the
  absorption and dispersion under four typical values of $\Delta$.}
  \label{Fig:modulationdelta}
\end{figure}

\begin{figure}[!htb]
  \centering
  \includegraphics[scale=0.5]{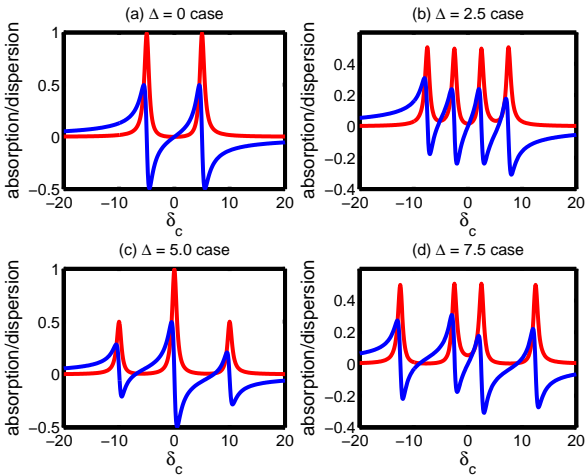}
  \caption{(color online) The absorption and dispersion when $\Delta$ = 0.0, 2.5, 5.0 \text{and} 7.5,
  respectively. Here, the blue curve represents the dispersion, while the red curve represents the absorption.
  The related parameters are $g_c = 5.0$, $\alpha = 0.001$, and $\beta = 0.666$. }
  \label{Fig:varydelta}
\end{figure}

Next, we fix the frequency of the probing light $\omega_p$, while scanning the Zeeman splitting
$\Delta$. The result is shown in Fig.\,\ref{Fig:modulationdelta}. It can be seen that when the
magnetic field is turned off, there is full absorption at $\delta_c = g_c$ [see
Fig.~\ref{Fig:varydelta}\,(a)]. This is understandable, since when $\Delta=0$ this system is
degenerated as a two-level configuration, and thus there exists only full absorption. After
applying the magnetic field, the absorption peaks are split into two halves, and the transparency
windows are established [Fig.~\ref{Fig:varydelta}\,(b)-(d)]. We want to further mention that
similar phenomena, such as the EIA, and the overlap of two transparency windows will happen in the
scan of the Zeeman splitting.

\section{Discussion and conclusion}
    The experimental realization using this four-level system has the following concerns:
    the real atomic system, the damping rates for each level, the magnitude of the magnetic field
    and the influence of atomic collisions.

    There have been many atomic systems that can satisfy our requirements, such as the
    4f$^6$6s$^2$~$^7F_1\leftrightarrow$~4f$^6$6s5p~$^6D_0$ transition of Sm\,\cite{smatom_1}-\cite{smatom_2},
    the 2p$^5$3s $^3P_1\leftrightarrow$~2p$^5$3p~$^3P_0$ transition of Ne\,\cite{neatom}, and also the
    5s$^1$ $^2S_{1/2}\leftrightarrow$ 5p$^1$ $^2P_{3/2}$ transition of $^{87}$Rb (D2 line).
    Since the laser for the wavelength of $\lambda = 780$~nm is commercially very
    popular, this adoption may be more readily for applications.

\begin{figure}[!htb]
  \centering
  \includegraphics[scale=0.4]{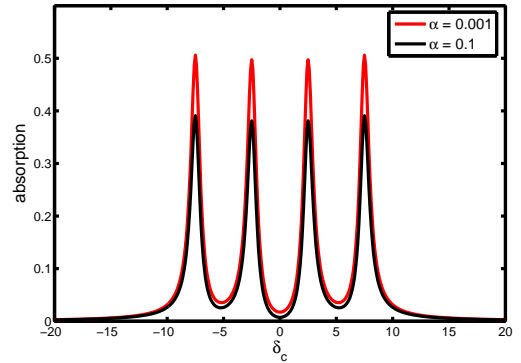}
  \caption{(color online) The impact of non-radiative damping. The red and black curve corresponds
  to a non-radiative damping rate of $\alpha = 0.001, 0.1$, respectively. Here, the related
  parameters are: $\beta = 0.666$, $\Delta = 2.5$, and $g_p = 5.0$.}
  \label{Fig:nonradiative}
\end{figure}

    Another important issue is the impact of damping rates. Since the dark state of this system
    [see Eq. \eqref{darkstate}] does not contain the upper level $\ket{1}$, it is immune to
    radiative damping. Therefore, this four-level tripod-type system is much more rigid,
    in fighting with decoherence due to radiative dampings, than a
    four-level $N$-type atomic system\,\cite{n-type}. Then, a question naturally arises: is the non-radiative
    damping, i.e., the dephasing from collisions, an urgent problem in the current configuration?
    This question is also crucial for the validity of the approximation used in Eq.~\eqref{formula}.
    Fig.~\ref{Fig:nonradiative} shows the absorption of the tripod-type system under
    different non-radiative damping rates, where $\alpha$ is the ratio of
    non-radiative to radiative damping rate, as we defined before. Fortunately,
    We can see that the general properties, i.e., the central frequency and width of the transparency
    windows, have been kept. Only the maximal absorption is reduced, which is not a crucial
    problem, since it still yields a good contrast. Moreover, given that non-radiative damping is
    much smaller than the radiative damping, its limitations on operation time is significantly
    less. Therefore, we can conclude that the non-radiative damping rates, though slightly
    changes the absorption of this
    system, is not a crucial problem, and our previous approximation is valid.

    Next, we briefly discuss the requirement of the magnetic field. Since the normal Zeeman splitting
    and the external magnetic field $B$ has the relation:
    $\Delta = g \mu_b B/\hbar$, where $g$ is the Land\'{e} factor and $\mu_b$
    is the Bohr magneton, then, if we need to realize a frequency shift in $^{87}$Rb of, say,
    10~MHz, a magnetic field of about 7\,Gauss is needed and can be easily fulfilled.

 In this paper, we discuss in detail the properties of a four-level tripod-type EIT system. We first
 show that this system has two transparency windows, then analysis are made to
 demonstrate that this configuration is actually a combination of two three-level $\Lambda$-type
 EIT systems. Following that, we focus on the discussions of the properties
 of the transparency windows.
 We show that the profile of the transparency windows rely strongly on the relative
 magnitude of the Rabi frequency of the coupling light and the Zeeman splitting.
 In the end, issues related to the experimental realization of this scheme are discussed.
 As a whole, it is pointed that the four-level tripod-type configuration yields an ideal dark
 state, which is very rare in four-level systems and is very beneficial for obtaining EIT in
 four-level systems. Besides this, it should also be noted that the ideal double dark states and
 double EIT windows have potential applications in the light storage for, at least,
 two frequencies of light and thus, could be connected with all-optical communication,
 together with the usage of wave division multiplexing (WDM) techniques.

\section*{Acknowledgments} The fruitful discussions with Bin Luo and Professor Anhong Dang are
greatly appreciated. This work is supported by the National Natural Science Foundation of China
(Grant No. 10474004), National Key Basic Research Program (Grant No. 2006CB921401) and DAAD
exchange program: D/05/06972 Projektbezogener Personenaustausch mit China (Germany/China Joint
Research Program).

\begin{IEEEbiographynophoto}{Xiao Li}
Born on March 26th, 1986, Fujian province, China. Xiao Li is now an undergraduate student and
studies at School of Earth and Space Science, Peking University. His current research interests
are: quantum coherence based on light-atom interactions, light storage and entanglement dynamics.
\end{IEEEbiographynophoto}
\begin{IEEEbiographynophoto}{Yu Liu}
Born on January 3rd, 1984, Tianjin, China. Yu Liu got Bachelor's degree in 2006 at Peking
University. He is now a PhD student at School of Electronics Engineering and Computer Science,
Peking University. His main research areas are: quantum coherence and interference and light
storage, quantum entanglement.
\end{IEEEbiographynophoto}
\begin{IEEEbiographynophoto}{Hong Guo}
Born on March 28th, 1969, Sichuan province, China. Hong Guo got Ph D in 1995 at Shanghai Institute
of Optics and Fine Mechanics, the Chinese Academy of Sciences. After that, he joined South China
Normal University for postdoctoral research work and in 1997 he got the full professorship. He is
now working as a full professor with School of Electronics Engineering and Computer Science, Peking
University. His main research areas are: quantum coherence and interference and light storage,
quantum entanglement, quantum cryptography and quantum key distribution, and laser propagation.
\end{IEEEbiographynophoto}
\clearpage
\end{document}